\documentclass[10pt,conference]{IEEEtran}
\usepackage{epsfig,setspace,amsmath,epsf,amssymb,bm,theorem,cite,graphicx,epstopdf,algorithm,float,color,mathtools,authblk,physics}
\usepackage[table,xcdraw]{xcolor}
\usepackage{subcaption}
\usepackage{booktabs}
\usepackage{tabularx}
\usepackage{xcolor}
\usepackage{algorithm,algorithmic}
\usepackage{dsfont}
\usepackage[short,c2]{optidef}

\newtheorem{theorem}{Theorem}

\newtheorem{definition}{Definition}
\newtheorem{remark}{Remark}
\newtheorem{lemma}{Lemma}

\newcommand{\st}{{\text{s.t.}}}
\newcommand{\e}{{\mathbb{E}}}

\IEEEoverridecommandlockouts
\allowdisplaybreaks

\begin{document}

\title{The Rate-Distortion-Deception Tradeoff}

\author[1]{Semih Akkoc}
\author[1]{Sahan Liyanaarachchi}
\author[1]{Sennur Ulukus}
\author[2]{Aylin Yener}

\affil[1]{\normalsize University of Maryland, College Park, MD, USA}
\affil[2]{\normalsize The Ohio State University, Columbus, OH, USA}

\maketitle

\begin{abstract}
    The problem of finding the optimal compression rate for a given random variable  has been traditionally studied under two main constraints: distortion and perception. The distortion constraint enforces the fidelity of our reconstruction with respect to the observed realization of the random variable, while the perception constraint ensures that the reconstruction is close to a sample from the distribution of the random variable of interest. In this work, we explore the possibility of reconstruction, such that the reconstructed sample is still within a desired fidelity level with our original realization of the random variable, but at the same time, it resembles a sample from a different \emph{target} distribution. We term this criterion as the \emph{deception} constraint and find the fundamental tradeoffs of rate-distortion and deception.
\end{abstract}

\section{Introduction}
Lossy source coding is a fundamental communication problem which follows from Shannon's seminal work in \cite{shannon}. This problem typically involves finding the theoretical limit of compression, i.e., the minimum number of bits required, to represent a random variable such that the reconstruction of the random variable generated using this compressed representation is within a desired level of fidelity from the original random variable. This fidelity constraint is often referred to as the distortion and is usually measured using a distance metric such as the mean squared error (MSE) or the Hamming distance. 

Since its formulation, several variations of this problem have been studied. \cite{wyner_side_info} characterizes the set of achievable rates with side information at the decoder. \cite{many_decoders} looks into optimal compression schemes with multiple decoders and \cite{cascaded_coding} studies the problem for cascaded communications systems whereas the work \cite{yener_relay} considers optimal compression schemes for relays with self source. Most of these works have studied the compression problem under only the distortion constraint. 

With recent advances in neural compression in the computer vision domain, a new dimension for optimal compression has been introduced: the distribution of the reconstruction must be similar to the distribution of original random variable. This idea was rooted in the fact that the reconstructed images must be appealing for human perception, and hence, reconstructing a sample from the original image distribution may be more valuable in this domain, even though it may have slightly more distortion. This new constraint on optimal compression was termed as the \emph{perception} constraint and was first formalized by Blau and Michaeli in \cite{blau_rdp}. 

In \cite{blau_rdp}, perception was defined as a distance metric between distributions. Following this definition, \cite{blau_rdp} introduced the rate-distortion-perception (RDP) function and found the rates that simultaneously satisfy the distortion and perception constraints. Later, \cite{wagner_achievabilty} showed that any compression rate which is greater than the RDP function is indeed achievable in the presence of infinite common randomness at the encoder and the decoder. Since then, several variations of this problem have been studied. For instance, \cite{wagner_finite} evaluates the role of common randomness and how the rate function varies in the presence of finite common randomness. This work was then extended in \cite{deniz_rdp_sideinfo} to incorporate side information into the problem. \cite{universal_rdp} studies the RDP function under Wasserstein distance and Kullback-Leibler (KL) divergence as the perception metrics and gives the explicit expression for scalar Gaussian compression.  \cite{vector_gauss_rdp} finds the RDP function for a vector Gaussian source and shows that the optimal reconstruction must be jointly Gaussian with the Gaussian source vector.

In this work, we generalize the concept of perception and consider the possibility of matching the distribution of the reconstruction to a \emph{target distribution}, instead of the default case of the source distribution. As we will detail shortly, for this formulation, we identify two major applications: deception and privacy. Here, we will go with the first application and refer to this criterion as the deception criterion. 
\begin{itemize}
\item \emph{Deception:} Consider a scenario where the receiver provides a service based on the received samples but would only continue providing the service if its estimate of the mean of the source random variable is lower than some threshold. The receiver may estimate the mean of the source random  variable through the received samples. In this scenario, an adversary may try to encode its data such that the distribution of the reconstruction is closer to a distribution with a low mean in order to continually receive service. But this would happen at the expense of the quality of the service received. Thus, an adversary may misrepresent its statistical characteristics, i.e., \emph{deceive} the receiver, in hopes of continually receiving service. 
\item \emph{Privacy:} Consider a scenario where we want to get a price estimate from a manufacturer based on a 3D point cloud model. In this case, we may prefer not to share our exact data with the manufacturer, but instead first find out the price of the closest model in their inventory that matches with our point clouds. This may be viewed as reconstructing a 3D point cloud that is close to our original point cloud but is from the manufacturer's distribution. By trying to statistically misrepresent our source random variable, this framework essentially enhances our \emph{privacy}.
\end{itemize}

Relevant to the notion of deception addressed in the first point above, reference \cite{yener_rdp_summarization} extends the rate-distortion framework for text summarization introduced in \cite{yener_rd_summarization} by incorporating perception constraints, for measuring perceptual quality of the summary with respect to the reference distribution instead of the original text distribution.

To summarize our contributions: In this paper, we formally define the concept of deception for optimal compression. We provide a rate-distortion-deception (RDD) function which completely characterizes the achievable rates under given distortion and deception constraints. Then, we evaluate this RDD function for several simple source distributions.

\section{Problem Formulation}
Let $X$ be the random variable that we are interested in compressing and let $\hat{X}$ be its reconstruction. Let $\mathcal{X}$ be the support of $X$. Let $d(x,\hat{x})$ be a distance metric which measures the distortion between our realization $x$ of $X$ and its reconstruction $\hat{x}$. Let $P_Y$ denote the target distribution for our reconstruction and $\mathcal{Y}$ be its support. For the deception metric, we will use the KL divergence. Our goal is to compress $X$ such that $\e[d(X,\hat{X})]\leq D$ and $D_{KL}(P_{\hat{X}}|| P_Y)\leq P$. The former constraint is known as the distortion constraint and the latter is termed as the deception constraint. Note that for the deception constraint to be valid, we require that the support of $\hat{X}$ is contained in $\mathcal{Y}$. For the design of our compression schemes, we will utilize stochastic encoders and decoders with the assumption that there is unlimited common randomness $U$ with support $\mathcal{U}$, available at the encoder and the decoder. Now, we define the achievable rate $R$ under these constraints.

\begin{definition}
    The triplet $(R,D,P)$ is achievable if $\forall \epsilon>0$, there exists an $n\in \mathbb{N}$, an encoding function $f:\mathcal{X}^n\times \mathcal{U}\to\mathbb{N}_0$ and a decoding function $g:\mathbb{N}_0\times \mathcal{U}\to \mathcal{Y}^n$, such that,
    \begin{align}
        g(f(X^n,U),U)&=\hat{X}^n,\\
        \frac{1}{n}\sum_{i=1}^n\e[d(X_i,\hat{X}_i)]&\leq D,\label{eqn:n_dist}\\
        \frac{1}{n}\sum_{i=1}^n D_{KL}(P_{\hat{X}_i}||P_Y)&\leq P,\label{eqn:n_deception}\\
        \frac{H(f(X^n,U)|U)}{n}&< R+\epsilon.\label{eqn:n_entropy}
    \end{align}
\end{definition}

Next, to characterize the achievable region of the triplets $(R,D,P)$, we define the information rate-distortion-deception function as follows,
\begin{align}
    R_Y(D,P)=\inf_{p_{\hat X| X}} & \ I(X;\hat{X})\nonumber\\
    \st \ & \ \e[d(X,\hat{X})]\leq D, \nonumber\\
    \quad  & \ D_{KL}(P_{\hat{X}}||P_Y)\leq P. \label{rdp_y}
\end{align}
In \eqref{rdp_y}, if $P_Y=P_X$, then $R_Y(D,P)$ is the rate-distortion-perception function $R_X(D,P)$ defined in \cite{blau_rdp}. 

\begin{remark}
    Note that, unlike the $R_X(D,P)$ function, the $R_Y(D,P)$ function is not well-defined for all possible combinations of $D$ and $P$. For instance, if $D_{KL}(P_X||P_Y)>0$, then we cannot simultaneously satisfy constraints for $D=0$ and $P=0$ in the case of rate-distortion-deception even with infinite rate, whereas in the case of rate-distortion-perception, the constraints $D=0$ and $P=0$ can be simultaneously achieved with infinite rate. Therefore, we define the $R_Y(D,P)$ function only at points for which the set $S_{D,P}=\{p_{\hat{X}|X}:\e[d(X,\hat{X})]\leq D, \ D_{KL}(P_{\hat{X}}||P_Y)\leq P\}$ is not empty.
\end{remark}

\begin{lemma}\label{lem:set_mono}
    For $D\leq D'$ and $P\leq P'$, if $S_{D,P}\neq \emptyset$ then $S_{D',P'}\neq \emptyset$.
\end{lemma}

\begin{lemma}\label{lem:set_convex}
     Let $S=\{(D,P): S_{D,P}\neq\emptyset\}$. Then, $S$ is a convex set.
\end{lemma}

\begin{lemma}\label{lem:RDP_convex}
    $R_Y(D,P)$ is convex and monotonically decreasing in each parameter in $S$.
\end{lemma}

The proof of Lemma \ref{lem:set_mono} follows from the fact that $S_{D,P}\subseteq S_{D',P'}$, and the proof of Lemma \ref{lem:set_convex} follows from the convexity of the KL divergence. The proof of Lemma \ref{lem:RDP_convex} is presented in Appendix \ref{appen:lem_RDP_convex}. Now, we formally define the achievable region.

\begin{theorem}\label{thrm:achievebility}
    The triplet $(R,D,P)$ is achievable if and only if $S_{D,P}\neq\emptyset$ and $R\geq R_Y(D,P)$.
\end{theorem}

The proof of Theorem \ref{thrm:achievebility} is given in Appendix \ref{appen:thrm_achievability}. In the next few sections, we explicitly derive the $R_Y(D,P)$ function for a few standard random variables.

\section{Bernoulli Source}
Here, we consider that both $P_X$ and $P_Y$ are Bernoulli distributions and use the Hamming distance as our distortion metric. Since the support of $\hat{X}\subseteq \mathcal{Y}$, $\hat{X}$ will also be Bernoulli. Let $p_{1|0}=\mathbb{P}(\hat{X}=1|X=0)$, $p_{1|1}=\mathbb{P}(\hat{X}=1|X=1)$, $p_{0|0}=1-p_{1|0}$ and $p_{0|1}=1-p_{1|1}$. Therefore, if $X\sim \text{Ber}(p)$, then $\hat{X}\sim \text{Ber}(q)$ where $ q=(1-p)p_{1|0}+p(1-p_{0|1})$. Now, the mutual information, distortion and deception criteria become,
\begin{align}
    &I(X; \hat{X}) = H_b(q) - (1-p)H_b(p_{1|0}) - p H_b(p_{0|1}),\\
    &\mathbb{E}[d(X, \hat{X})] = \mathbb{P}(X \neq \hat{X}) = (1-p)p_{1|0} + p p_{0|1},\\
    &D_{KL}(P_{\hat{X}} || P_Y) = (1-q) \ln \frac{1-q}{P_Y(0)} + q \ln \frac{q}{P_Y(1)}, 
\end{align}
where $H_b(x) = -x \ln x - (1-x) \ln(1-x)$.  Then, for any $(D,P)\in S$, we can solve the $R_Y(D,P)$ function by solving the following optimization problem,
\begin{align}
    \inf_{p_{1|0},p_{0|1}\in[0,1]} & \ H_b(q) - (1-p)H_b(p_{1|0}) - p H_b(p_{0|1})\nonumber\\
    \st \qquad & \ (1-p)p_{1|0} + p p_{0|1}\leq D,\nonumber\\
    & \  (1-q) \ln \frac{1-q}{P_Y(0)} + q \ln \frac{q}{P_Y(1)} \leq P.
\end{align}

We solve the above optimization problem using a Lagrangian formulation. Let $\lambda \ge 0$ and $\beta \ge 0$ be the Lagrange multipliers for the distortion constraint and for the deception constraint, respectively. Then, the Lagrangian $\mathcal{L}(p_{\hat{x}|x},\lambda,\beta)$ is,
\begin{align}
    \mathcal{L}(p_{\hat{x}|x},\lambda,\beta) &= H_b(q) - (1-p)H_b(p_{1|0}) - p H_b(p_{0|1}) \nonumber \\
    &\quad + \lambda \Big( (1-p)p_{1|0} + p p_{0|1} - D \Big) \nonumber \\
    &\quad + \beta \left( (1-q) \ln \frac{1-q}{P_Y(0)} + q \ln \frac{q}{P_Y(1)} - P \right). 
\end{align}

Note that the above optimization problem is convex and strictly feasible for $(D,P)\in S$. Hence, the KKT conditions yield necessary and sufficient conditions for optimality. Therefore, at the optimal point, we have,
\begin{align}
    \ln \frac{p_{1|0}}{1-p_{1|0}} &= \ln \frac{q}{1-q} - \lambda - \beta \ln \left( \frac{q}{1-q} \frac{P_Y(0)}{P_Y(1)} \right), \\
    \ln \frac{p_{0|1}}{1-p_{0|1}} &= \ln \frac{1-q}{q} - \lambda + \beta \ln \left( \frac{q}{1-q} \frac{P_Y(0)}{P_Y(1)} \right). 
\end{align}
Now, we can find $p_{1|0}$ and $p_{0|1}$ in terms of $q$ as follows,
\begin{align}
    p_{1|0} = \frac{\rho e^{-\lambda}}{1 + \rho e^{-\lambda}} , \quad
    p_{0|1} = \frac{\rho^{-1} e^{-\lambda}}{1 + \rho^{-1} e^{-\lambda}},
\end{align}
where $\rho \triangleq \left(\frac{q}{1-q}\right)^{1-\beta} \left(\frac{P_Y(1)}{P_Y(0)}\right)^\beta$. Note that since the above expressions for $p_{1|0}$ and $p_{0|1}$ are naturally in the region $[0,1]$, we need not explicitly enforce the constraints $p_{0|1},p_{1|0}\in[0,1]$ in the Lagrangian. 

Now, substituting the above probabilities into $q=(1-p)p_{1|0}+p(1-p_{0|1})$ yields the following fixed point equation,
\begin{align}
    q = (1-p) \frac{\rho e^{-\lambda}}{1 + \rho e^{-\lambda}} + p \frac{1}{1 + \rho^{-1} e^{-\lambda}}. \label{eqn:fixed_point_q}
\end{align}

Next, we identify solutions in various regions based on the active constraints and complementary slackness conditions. For this, let $q^*$, $p_{1|0}^*$ and $p_{0|1}^*$ be the optimal solution to the optimization problem.

\subsubsection{Region 1. Classical Rate-Distortion}
In this regime, $\lambda > 0$ and $\beta = 0$. The deception constraint $P$ is loose and the distortion constraint is tight. In this case,  $\rho = \frac{q^*}{1-q^*}$ and the optimal solution is given as,
\begin{align}
    p_{1|0}^* = \frac{q^* e^{-\lambda}}{1 - q^* + q^* e^{-\lambda}}, \quad p_{0|1}^* = \frac{(1-q^*) e^{-\lambda}}{q^* + (1-q^*) e^{-\lambda}}. 
\end{align}
Now, $e^{-\lambda}$ can be found in terms of $q^*$ using  $\mathbb{E}[d(X,\hat{X})] = D$ which can be substituted into the fixed point equation to find  $q^*$.

\subsubsection{Region 2. Zero-Rate Regime}
When $\lambda = 0$ and $\beta \geq 0$, the distortion constraint is loose. This implies $e^{-\lambda} = 1$, and
\begin{align}
    p_{1|0}^* = \frac{\rho}{1+\rho}, \quad p_{0|1}^* = \frac{1}{1+\rho}. 
\end{align}
Notice that $\mathbb{P}(\hat{X}=1|X=0)= p_{1|0}^* = 1 - p_{0|1}^* = \mathbb{P}(\hat{X}=1|X=1)$. Hence, the output $\hat{X}$ can be generated independently of $X$, and the rate falls to $R_Y(D,P) = 0$. $\hat{X}$ can be independently generated as long as $D\geq \min_{q'\in[q_L(P), q_U(P)]}(p+q'-2pq')$ where  $q_L(P)$ and $q_U(P)$ are the roots of $D_{KL}(q||P_Y)=P$.

\subsubsection{Region 3. The RDD Tradeoff Regime}
Here, both constraints are tight ($\lambda > 0$, $\beta > 0$). Using $\e[d(X,\hat{X})]=D$,
\begin{align}
    p^*_{1|0}&=\frac{D+q^*-p}{2(1-p)},\\
    p^*_{0|1}&=\frac{D-q^*+p}{2p}.
\end{align} 
Therefore, when the distortion constraint is active, the objective function will be convex with respect to $q$. Let $q_{RD}(D)$ be the optimal $q^*$ in the absence of the deception constraint. We will operate in this Region 3 only if $q_{RD}(D)\notin[q_L(P),q_U(P)]$. Since the objective function is convex when the distortion constraint is active, if $q_{RD}(D)<q_L(P)$, then $q^*=q_L(P)$ and if $q_{RD}(D)>q_U(P)$, then $q^*=q_U(P)$. 

Further, since $p_{0|1},p_{1|0}\in[0,1]$, we will operate in this region, only if we get $D\geq \min\{|p-q_L(P)|,|p-q_U(P)|\}$. 

\begin{figure}
    \centering
    \includegraphics[width=\linewidth]{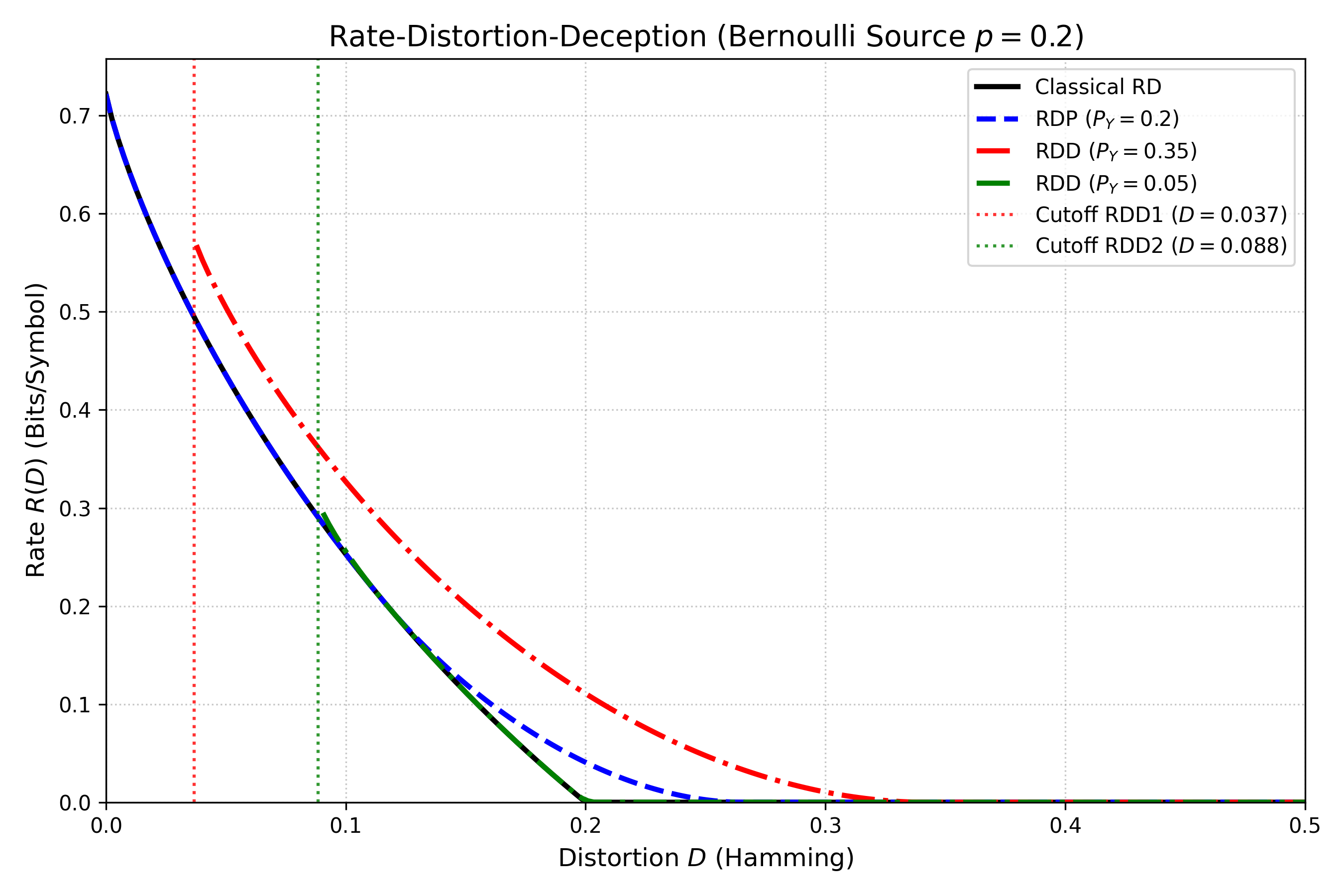}
    \caption{Comparison of the optimal compression rates with perception and deception for a Bernoulli source and Bernoulli target distribution.}
    \label{fig:rdp_vs_rdd}
\end{figure}

\begin{remark}
To illustrate the difference between RDP and RDD functions, we plot the rate-distortion curves with fixed perception and deception constraints. As seen in Fig.~\ref{fig:rdp_vs_rdd}, unlike the RDP curve (blue), the RDD curves do not span across every possible distortion. It has a sharp cutoff beyond which attaining that particular distortion level is not possible even with infinite rate. Moreover, we see that depending on the target distribution, RDD curve may lie above (red curve) or below (green curve) the RDP curve (blue curve). This is due to the fact that when the target distribution is close to the reconstruction with only the distortion constraint, we can satisfy the deception constraint without any additional rate.
\end{remark}

\section{Bernoulli Source -- Ternary Reconstruction}
For this case we consider that $P_X$ has Bernoulli distribution but $P_Y$ has ternary distribution with probability mass function $\{y_0,y_1,y_2\}$ on the support $\mathcal{Y}=\{0,1,2\}$. Again, we use Hamming distance as our distortion metric. $\hat{X}$ will be ternary since support of $\hat{X} \subseteq \mathcal{Y}$. 

We will express the mutual information, distortion and deception constraints explicitly as sums over the discrete alphabets $\mathcal{X} = \{0, 1\}$ and $\hat{\mathcal{X}} = \mathcal{Y} = \{0, 1, 2\}$ to form the Lagrangian. Note that, $P_{\hat{X}}$ can be expressed using  $p_{\hat{x}|x}$ as,
\begin{align}
    P_{\hat{X}}(\hat{x}) = \sum_{x=0}^{1} P_X(x) p_{\hat{x}|x}. 
\end{align}
Now, the optimization problem at hand can be expressed as,
\begin{align}
    \inf_{p_{\hat{x}|x}}& \quad I(X;\hat{X}) \nonumber\\
    \st & \quad \mathbb{E}[d(X,\hat{X})] \le D, \nonumber \\
    &\quad D_{KL}(P_{\hat{X}} || P_Y) \le P, \nonumber \\
    &\quad \sum_{\hat{x}=0}^{2} p_{\hat{x}|x} = 1, \ \forall x. 
\end{align}

Let $\lambda \ge 0$ and  $\beta \ge 0$ represent the Lagrange multipliers for the distortion and the deception constraints, respectively. Let $\gamma_x$ denote the Lagrange multiplier for probability simplex constraints. Then, the Lagrangian $\mathcal{L}(p_{\hat{x}|x},\lambda,\beta,\gamma)$ is,
\begin{align}
    \mathcal{L}(p_{\hat{x}|x},\lambda,\beta,\gamma_x) &= I(X;\hat{X}) + \lambda \left( \mathbb{E}[d(X,\hat{X})] - D \right) \nonumber \\
    &\quad + \beta \left( D_{KL}(P_{\hat{X}} || P_Y) - P \right) \nonumber \\
    &\quad + \sum_{x=0}^{1} \gamma_x \left( \sum_{\hat{x}=0}^{2} p_{\hat{x}|x} - 1 \right). 
\end{align}
Expanding this function yields,
\begin{align}
    \mathcal{L}(p_{\hat{x}|x},\lambda,\beta,\gamma_x) &= \sum_{x=0}^{1} \sum_{\hat{x}=0}^{2} P_X(x) p_{\hat{x}|x} \log \frac{p_{\hat{x}|x}}{P_{\hat{X}}(\hat{x})} \nonumber \\
    &\quad + \lambda \left( \sum_{x=0}^{1} \sum_{\hat{x}=0}^{2} P_X(x) p_{\hat{x}|x} d(x,\hat{x}) - D \right) \nonumber \\
    &\quad + \beta \left( \sum_{\hat{x}=0}^{2} P_{\hat{X}}(\hat{x}) \log \frac{P_{\hat{X}}(\hat{x})}{y_{\hat{x}}} - P \right) \nonumber \\
    &\quad + \sum_{x=0}^{1} \gamma_x \left( \sum_{\hat{x}=0}^{2} p_{\hat{x}|x} - 1 \right). 
\end{align}

Note that the above optimization problem is convex and strictly feasible for $(D,P)\in S$. The KKT conditions yield the following condition for optimality, when $P_X(x)\neq0$,
\begin{align}
    \log \frac{p_{\hat{x}|x}}{P_{\hat{X}}(\hat{x})} + \lambda d(x,\hat{x}) + \beta \left( \log \frac{P_{\hat{X}}(\hat{x})}{y_{\hat{x}}} + 1 \right) + \frac{\gamma_x}{P_X(x)} = 0. 
    \label{eqn:kkt_ternary}
\end{align}
We isolate $\log p_{\hat{x}|x}$ and exponentiating \eqref{eqn:kkt_ternary} to obtain,
\begin{align}
    p^*_{\hat{x}|x} = \frac{ P^*_{\hat{X}}(\hat{x})^{1-\beta} \cdot y_{\hat{x}}^{\beta} \cdot \exp(-\lambda d(x,\hat{x})) }{ \sum_{i=0}^{2} P^*_{\hat{X}}(i)^{1-\beta} \cdot y_{i}^{\beta} \cdot \exp(-\lambda d(x,i))}. 
\end{align}
For $\beta = 0$, we obtain the standard rate-distortion solution. If $\beta = 1$, then $p_{\hat{x}|x}^*$ depends strictly on the target distribution $y_{\hat{x}}$. $p^*_{\hat{x}|x}$ does not have a closed-form solution since the solution for $p^*_{\hat{x}|x}$ functionally depends on $P_{\hat{X}}(\hat{x})$, which in turn, depends circularly on $p^*_{\hat{x}|x}$. Thus, in order to numerically evaluate $p^*_{\hat{x}|x}$, we use the Blahut-Arimoto (BA) algorithm. 

The BA algorithm starts with an initialization step, where we choose a valid initial output distribution $Q^{(0)}(\hat{x}) > 0$ for $\hat{x} \in \{0,1,2\}$ (such as $Q^{(0)} = P_Y$). Then, we select fixed $(\lambda,\beta)$ pair where $\lambda > 0$ and $1 \ge \beta \ge 0$. In Step 1, given the current output marginal estimate $Q^{(t)}$, the transition matrix is updated using the derived solution for all $x \in \{0, 1\}$,
\begin{align}
    p_{\hat{x}|x}^{(t+1)} = \frac{ \left( Q^{(t)}(\hat{x}) \right)^{1-\beta} \cdot y_{\hat{x}}^{\beta} \cdot \exp(-\lambda d(x,\hat{x})) }{ \sum_{i=0}^{2} \left( Q^{(t)}(i) \right)^{1-\beta} \cdot y_{i}^{\beta} \cdot \exp(-\lambda d(x,i))}. 
\end{align}
In Step 2, the auxiliary output distribution is updated by calculating the marginal and using the source distribution,
\begin{align}
    Q^{(t+1)}(\hat{x}) & = P_X(0) p_{\hat{x}|0}^{(t+1)} + P_X(1) p_{\hat{x}|1}^{(t+1)} \\
    & = \sum_{x=0}^{1} P_X(x) p_{\hat{x}|x}^{(t+1)}.
\end{align}

\begin{figure*}[!t] 
    \centering
    \begin{subfigure}[b]{.305\textwidth}
        \centering
        \includegraphics[trim={1.5cm 1.5cm 1.5cm 2cm}, clip,width=\linewidth]{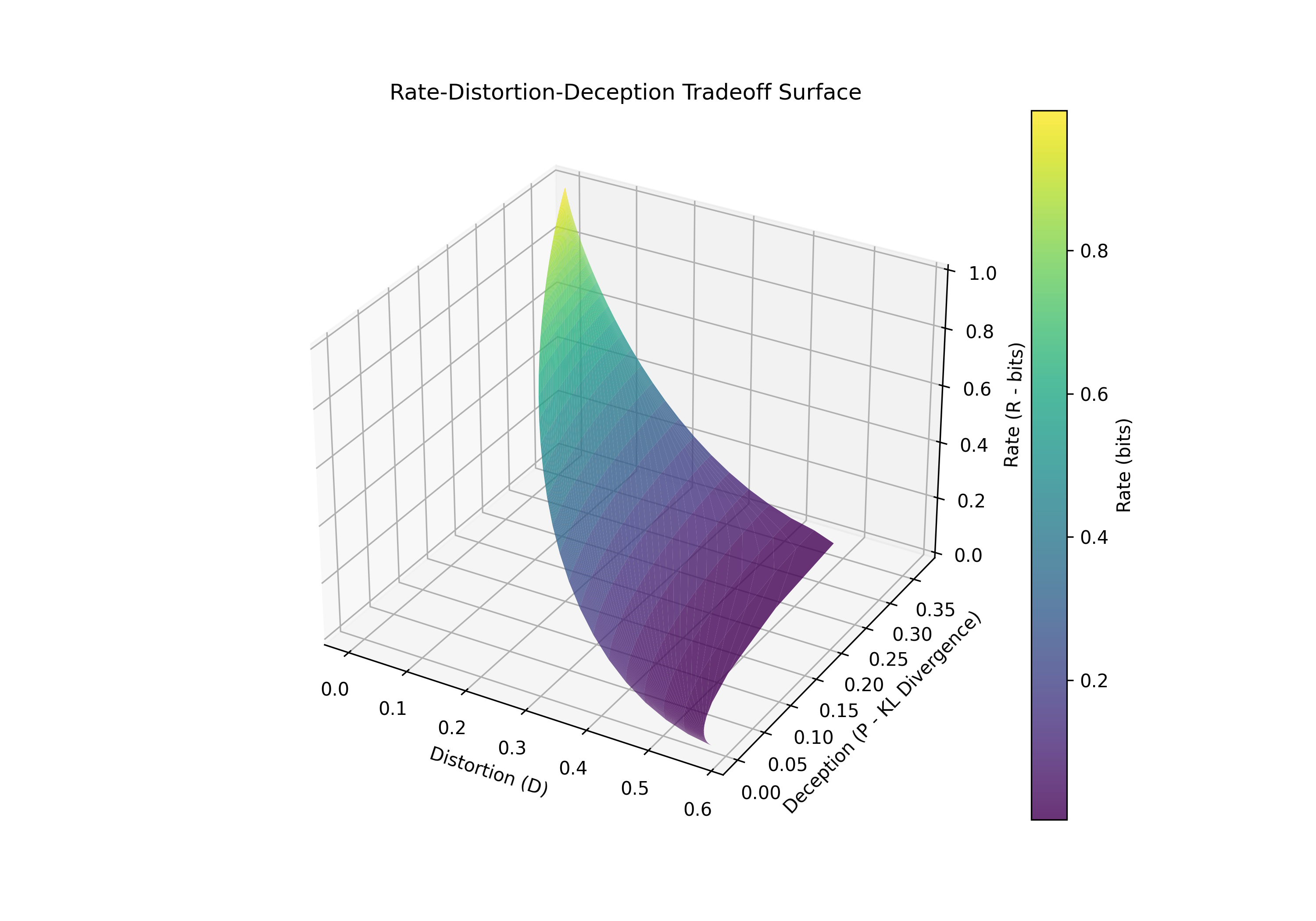}
        \caption{RDD surface}
        \label{subfig:rdp}
    \end{subfigure}%
    \hfill
    \begin{subfigure}[b]{.245\textwidth}
        \centering
        \includegraphics[trim={5cm 4cm 7cm 5cm}, clip, width=\linewidth]{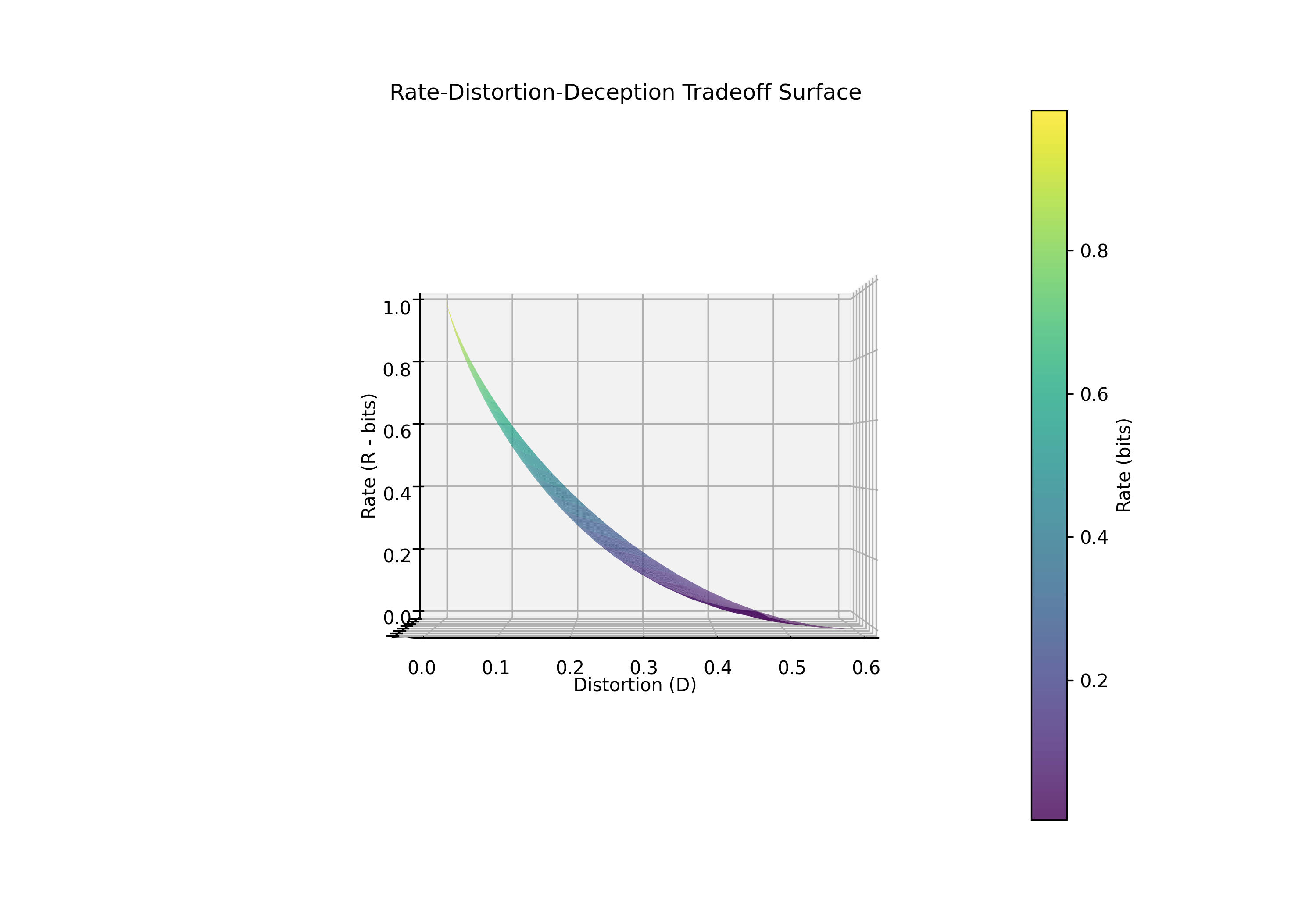}
        \caption{Rate-Distortion}
        \label{subfig:rd}
    \end{subfigure}%
    \hfill 
    \begin{subfigure}[b]{.245\textwidth}
        \centering
        \includegraphics[trim={5cm 4cm 7cm 5cm}, clip, width=\linewidth]{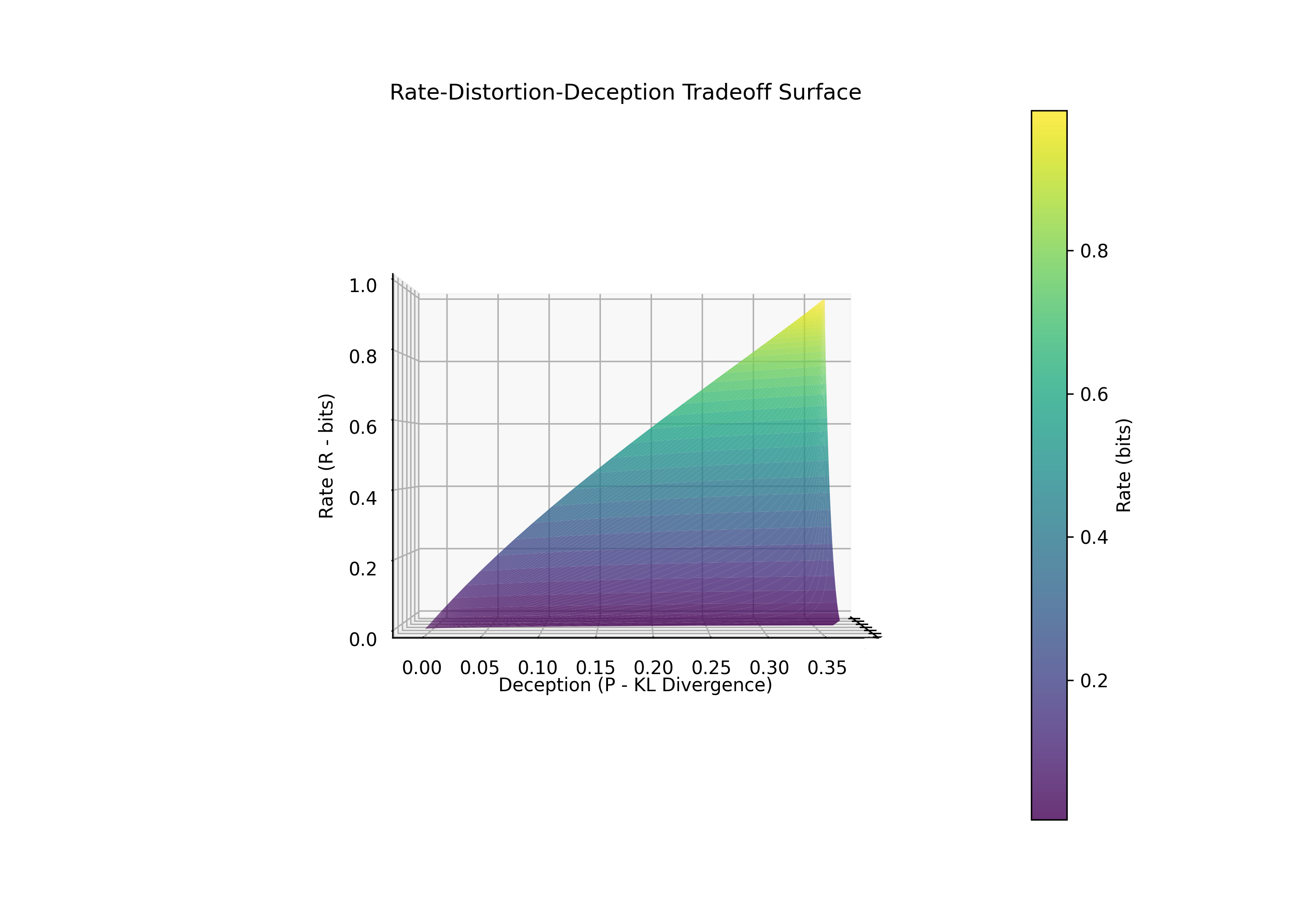}
        \caption{Rate-Deception}
        \label{subfig:rp}
    \end{subfigure}%
    \hfill 
    \begin{subfigure}[b]{.205\textwidth}
        \centering
        \includegraphics[trim={5cm 3.5cm 7cm 4.5cm}, clip, width=\linewidth]{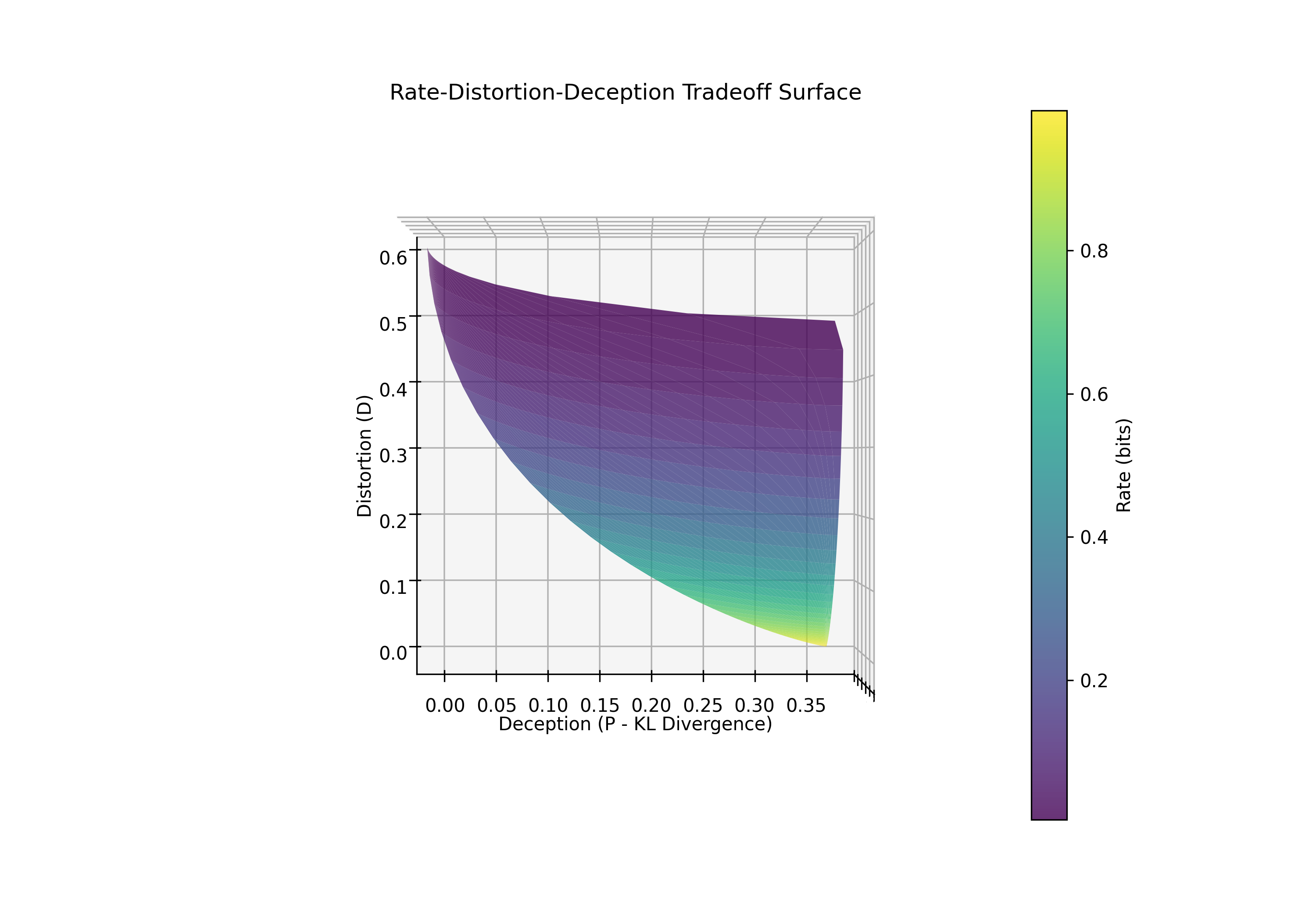}
        \caption{Distortion-Deception}
        \label{subfig:dp}
    \end{subfigure}
    
    \caption{Simulation results showing RDD tradeoff surface: (a) For $P_X=[0.5, 0.5]$ and $P_Y=[0.3,0.5,0.2]$. The points represent the convergence of the modified BA algorithm across a grid of $(\lambda,\beta)$. The conditional distribution $P_{\hat{x}|x}$ must map $X\in \{0,1\}$ to an output $\hat{X} \in\{0,1,2\}$. Since the Hamming distance is used, the state $\hat{x}=2$ acts entirely as a deception term making $\hat{x}=2$ no benefit for distortion meaning the $p_{\hat{x}|x}$ only uses this transition when forced by the deception multiplier $\beta$. As shown in (d), achieving the perfect deception $P\to0$ requires $P_{\hat{X}}$ to perfectly match $P_Y$. This forces the algorithm to push $20\%$ of data to the $\hat{x}=2$ creating a non-zero distortion of $D\geq0.2$ regardless of the rate. Further, (b) and (c) highlight the convex limits of the system. As $\beta \to 1$, $p_{\hat{x}|x}^*$ becomes more dependent on the target distribution $y_{\hat{x}}$ rather than the input $X$. This minimizes KL divergence at the expense of distortion, pushing the operational bounds toward the lower right part of the 3D surface.}
    \label{fig:rdp-all-ternary}
\end{figure*}

This numerical process repeats Steps 1 and 2 iteratively until it reaches convergence, i.e., when the maximum absolute difference between $Q^{(t+1)}$ and $Q^{(t)}$ drops below a predefined tolerance threshold $\epsilon$. To map out the complete $R_Y(D,P)$ operational surface, we compute this BA algorithm across a 2D grid of Lagrangian multipliers $(\lambda, \beta)$, calculating the resulting information rate $R$, expected distortion $D$, and deception $P$ at convergence for every $(\lambda, \beta)$ pair. The RDD function obtained using the approach is illustrated in Fig.~\ref{fig:rdp-all-ternary}.

\section{Scalar Gaussian Source}
Here, we consider that the source is $X \sim \mathcal{N}(\mu_X, \sigma_X^2)$, and the deception distribution is $Y \sim \mathcal{N}(\mu_Y, \sigma_Y^2)$. As distortion and deception metrics, we use MSE and KL-divergence, respectively. Similar to \cite{vector_gauss_rdp}, we show that the optimal reconstruction $\hat{X}$ should be jointly Gaussian with $X$. Thus, we can represent the reconstruction  as $\hat{X} \sim \mathcal{N}(\mu_{\hat{X}}, \sigma_{\hat{X}}^2)$. Since $\hat{X}$ is jointly Gaussian with $X$, we have that $\e[\hat{X}|X]$ is a linear function of $X$. Moreover, from the orthogonality principle, we have that $\hat{X}-\e[\hat{X}|X]$ is independent of $\e[\hat{X}|X]$. Therefore, we can express our estimator as,
\begin{align}
    \hat{X} &= \e[\hat{X}|X]+\hat{X}-\e[\hat{X}|X]\\
    &=aX +  Z+b,
\end{align}
where $Z \sim \mathcal{N}(0, \sigma_Z^2)$ is independent of $X$ and $a$, $b$ are constants. Then, the expected value of the reconstruction becomes $\mu_{\hat{X}} = a\mu_X + b$, and the variance of the reconstruction is written as $\sigma_{\hat{X}}^2 = a^2\sigma_X^2 + \sigma_Z^2$.

Now, we can write the mutual information, distortion and deception explicitly as,
\begin{align}
    I(X; \hat{X}) &= \frac{1}{2} \ln\left( \frac{a^2\sigma_X^2 + \sigma_Z^2}{\sigma_Z^2} \right), \\
    \mathbb{E}[(X-\hat{X})^2] & = (1-a)^2\sigma_X^2 + (\mu_X - \mu_{\hat{X}})^2 + \sigma_Z^2,  \\
    \label{eqn:gauss_dist}
    D_{KL}(P_{\hat{X}} || P_Y)&= \frac{1}{2}\bigg[ \ln\left(\frac{\sigma_Y^2}{a^2\sigma_X^2 + \sigma_Z^2}\right) \nonumber \\
    &\quad + \frac{a^2\sigma_X^2 + \sigma_Z^2 + (\mu_{\hat{X}} - \mu_Y)^2}{\sigma_Y^2} - 1 \bigg]. 
\end{align}

The Lagrangian $\mathcal{L}$ with Lagrange multipliers $\lambda \ge 0$ for the distortion constraint and $\beta \ge 0$ for the deception constraint is,
\begin{align}
    \mathcal{L}(a,\mu_{\hat{X}},&\sigma^2_Z,\lambda,\beta)= \frac{1}{2}\ln(a^2\sigma_X^2 + \sigma_Z^2) - \frac{1}{2}\ln(\sigma_Z^2) \nonumber \\
    & + \lambda\left[ (1-a)^2\sigma_X^2 + (\mu_X - \mu_{\hat{X}})^2 + \sigma_Z^2 - D \right] \nonumber \\
    & + \beta\bigg[ \frac{1}{2}\ln\left(\frac{\sigma_Y^2}{a^2\sigma_X^2 + \sigma_Z^2}\right) \nonumber \\
    & + \frac{a^2\sigma_X^2 + \sigma_Z^2 + (\mu_{\hat{X}} - \mu_Y)^2}{2\sigma_Y^2} - \frac{1}{2} - P \bigg]. 
\end{align}
Now, the KKT conditions yield the following,
\begin{align} 
    \mu_{\hat{X}} &= \frac{2\lambda \sigma_Y^2 \mu_X + \beta \mu_Y}{2\lambda \sigma_Y^2 + \beta}, \label{eqn:gauss_a} \\
    \frac{1}{2\sigma_Z^2} &= \frac{1-\beta}{2\sigma_{\hat{X}}^2} + \lambda + \frac{\beta}{2\sigma_Y^2}, \\ \label{eqn:gauss_sigma}
    a &= 2\lambda \sigma_Z^2.
\end{align}

Next, we investigate the active regions in this setting.

\subsubsection{Region 1. Classical Rate-Distortion}
In this region, we have $\lambda > 0$ and $\beta = 0$. Since $\beta = 0$, the mean equation gives $ \mu_{\hat{X}} = \mu_X $. Using $ a = 2\lambda\sigma_Z^2 $ and \eqref{eqn:gauss_sigma}, we get $\sigma_{\hat{X}}^2=a\sigma^2_X$. This yields the following relations 
\begin{align}
a = 1 - \frac{D}{\sigma_X^2}, \quad \sigma_Z^2 = aD, \quad \sigma_{\hat{X}}^2 = \sigma_X^2 - D, 
\end{align}
giving classical rate-distortion function $R_Y(D) = \frac{1}{2} \ln\left(\frac{\sigma_X^2}{D}\right)$.

\subsubsection{Region 2. Zero-Rate Regime}
When $\lambda=0$, we have $a=0$. This means that our reconstruction $\hat{X}$ can be generated independently of $X$, as long as there exist $\mu_{\hat{X}}$ and $\sigma^2_{\hat{X}}$ that satisfy the constraints $\sigma^2_X+(\mu_X-\mu_{\hat{X}})^2+\sigma_{\hat{X}}^2\leq D$ and $D_{KL}(\mathcal{N}(\mu_{\hat{X}},\sigma^2_{\hat{X}})||\mathcal{N}(\mu_{Y},\sigma^2_{Y}))\leq P$. If further $\beta>0$, this forces $\mu_{\hat{X}}=\mu_Y$ and $\sigma_{\hat{X}}^2=\sigma^2_Y$. However, when $\beta>0$, the deception constraint is tight. Thus, this scenario arises only when $P=0$. In either case, since $\hat{X}$ is independent of $X$, we have $R_Y(D)=0$. 

\subsubsection{Region 3. The RDD Tradeoff Regime}
When both constraints are active (i.e., $\lambda>0$, $\beta>0$), closed-form solutions are not available. However, when $\mu_Y=\mu_X$, the expressions can be simplified considerably. In this case, $\mu_{\hat{X}}=\mu_X=\mu_Y$. Thus, we can find $\sigma_{\hat{X}}$ and $a$ as the solutions of,
\begin{align}
    \sigma^2_{\hat{X}}-\sigma^2_Y\ln(\sigma^2_{\hat{X}})&=(2P+1-\ln(\sigma^2_Y))\sigma^2_Y,\\
    \sigma^2_X-2a+\sigma^2_{\hat{X}}&=D.
\end{align}

We will operate in Region 1 as long as $(D,P)$ satisfies,
\begin{align}
    P \ge D_{KL}\left(\mathcal{N}(\mu_X, \sigma_X^2 - D) || \mathcal{N}(\mu_Y, \sigma_Y^2)\right). 
\end{align}

We will operate in Region 2 if $ D \ge D_{min}(P) $, where $ D_{min}(P) $ is the minimum  MSE achievable, when $\hat{X}$ is generated independently, for a given deception budget. Setting $ a = 0 $, we have $ D = \sigma_X^2 + \sigma_{\hat{X}}^2 + (\mu_X - \mu_{\hat{X}})^2 $. Therefore, $D_{min}(P)$ can be expressed as
\begin{align}
    D_{min}(P)&= \min_{\sigma_{\hat{X}}^2, \mu_{\hat{X}}} \left( \sigma_X^2 + \sigma_{\hat{X}}^2 + (\mu_X - \mu_{\hat{X}})^2 \right) \text{,}\\
    \st\quad
    &\frac{1}{2}\left[ \ln\left(\frac{\sigma_Y^2}{\sigma_{\hat{X}}^2}\right) + \frac{\sigma_{\hat{X}}^2 + (\mu_{\hat{X}} - \mu_Y)^2}{\sigma_Y^2} - 1 \right] \le P. 
\end{align}

\section{Vector Gaussian Source}
In this section, we extend the $R_Y(D,P)$ to the case of vector Gaussian source. For this setup, we have source $X \sim \mathcal{N}(0, \Sigma_X)$ and deception distribution $Y \sim \mathcal{N}(0, \Sigma_Y)$, which are of dimension $M$. We use MSE and KL divergence for the distortion and deception metrics. Similar to \cite{vector_gauss_rdp}, we show that the optimal reconstruction is jointly Gaussian with $X$. Therefore, let  $\hat{X} \sim \mathcal{N}(0, \Sigma_{\hat{X}})$.

Let $A$ represent the conditional covariance of $\hat{X}$ given $X$,
\begin{align}
    A= \Sigma_{\hat{X}|X} = \Sigma_{\hat{X}} - \Sigma_{X\hat{X}}^T \Sigma_X^{-1} \Sigma_{X\hat{X}}.
\end{align}
We can write mutual information, distortion and deception as,
\begin{align}
    I(X;\hat{X}) &= \frac{1}{2} \ln |\Sigma_{\hat{X}}| - \frac{1}{2} \ln |A|, \\
    \mathbb{E}[||X-\hat{X}||^2] &= \text{tr}(\Sigma_X + \Sigma_{\hat{X}} - 2\Sigma_{X\hat{X}}), \\
    D_{KL}(P_{\hat{X}} || P_Y) &= \frac{1}{2} \Big( \text{tr}(\Sigma_Y^{-1} \Sigma_{\hat{X}}) - M + \ln \frac{|\Sigma_Y|}{|\Sigma_{\hat{X}}|} \Big).
\end{align}
Combining these terms and letting $\lambda\geq0$ and $\beta\geq0$ be the Lagrange multipliers for distortion and deception terms, respectively, we construct the Lagrangian as,
\begin{align}
    \mathcal{L}(\Sigma_{\hat{X}},\Sigma_{X\hat{X}},&\lambda,\beta) = \frac{1}{2} \ln |\Sigma_{\hat{X}}| - \frac{1}{2} \ln |\Sigma_{\hat{X}} - \Sigma_{X\hat{X}}^T \Sigma_X^{-1} \Sigma_{X\hat{X}}| \nonumber \\
    &+ \lambda \left( \text{tr}(\Sigma_X + \Sigma_{\hat{X}} - 2\Sigma_{X\hat{X}}) - D \right) \nonumber \\
    &+ \frac{\beta}{2} \Big( \text{tr}(\Sigma_Y^{-1} \Sigma_{\hat{X}}) -M  + \ln \frac{|\Sigma_Y|}{|\Sigma_{\hat{X}}|}-P \Big). 
    \label{eqn:lagrangian_vec_gauss}
\end{align}

We obtain the KKT conditions by differentiating $\mathcal{L}$ with respect to the $\Sigma_{X\hat{X}}$ and $\Sigma_{\hat{X}}$ as,
\begin{align}
    \frac{1}{2} \Sigma_{\hat{X}}^{-1} + \lambda I + \frac{\beta}{2} \Sigma_Y^{-1} &=  \frac{1}{2} A^{-1} + \frac{\beta}{2} \Sigma_{\hat{X}}^{-1}, \label{eqn:vecgauss_der2}\\
    \Sigma_{X\hat{X}} &= 2\lambda \Sigma_X A. \label{eqn:vecgauss_der1}
\end{align}
By substituting $A^{-1} = 2\lambda \Sigma_{X\hat{X}}^{-1} \Sigma_X$ from \eqref{eqn:vecgauss_der1} into the \eqref{eqn:vecgauss_der2}, we obtain the equation defining the optimal covariances,
\begin{align}
    -\lambda \Sigma_{X\hat{X}}^{-1} \Sigma_X + \frac{1-\beta}{2} \Sigma_{\hat{X}}^{-1} + \lambda I + \frac{\beta}{2} \Sigma_Y^{-1} = 0. 
    \label{gaussvec1}
\end{align}

We investigate the active regions based on the Lagrange multipliers $ \lambda \geq 0$ for distortion, and $ \beta \geq 0 $ for deception.

\subsubsection{Region 1. Classical Rate-Distortion}
In this region, the distortion constraint is active ($\lambda > 0$), while the deception constraint remains inactive ($\beta = 0$). The KKT equations simplify to standard \emph{reverse water-filling}. Then, the optimal reconstruction can be obtained by decomposing $X$ as $ X = \hat{X} + Z $, where $ Z \perp \hat{X} $. This implies $ \Sigma_{X\hat{X}} = \Sigma_{\hat{X}} $, and $A= \Sigma_{\hat{X}} - \Sigma_{\hat{X}} \Sigma_X^{-1} \Sigma_{\hat{X}} $. The solution is obtained by diagonalizing $ \Sigma_X = U \Lambda_X U^T $, then $ \Sigma_{\hat{X}} = U \max(0, \Lambda_X - \theta I) U^T $. The parameter $ \theta $ ensures that the trace constraint $ \text{tr}(\min(\Lambda_X, \theta I)) = D $ holds.

\subsubsection{Region 2. Zero-Rate Regime}
In here, $\hat{X}$ is independent of $X$. As long as there exists a $\Sigma_{\hat{X}}$ such that $\text{tr}(\Sigma_X + \Sigma_{\hat{X}})\leq D$ and $D_{KL}(\mathcal{N}(0,\Sigma_{\hat{X}})||\mathcal{N}(0,\Sigma_Y))\leq P$, we will be operating in this regime. When these conditions are satisfied, the optimal reconstruction becomes $\Sigma_{\hat{X}}=0$.

\subsubsection{Region 3. The RDD Tradeoff Regime}
When both constraints are active ($\lambda > 0$, $\beta > 0$), the system does not have a closed-form diagonalization unless $ \Sigma_X $ and $ \Sigma_Y $ are simultaneously diagonalizable (see Appendix \ref{appen:sim_diag_Gauss}). The definitive solution is instead found by numerically solving the equation derived in the KKT step:
\begin{align}
\Sigma_{\hat{X}} &= \left( \frac{1-\beta}{2\lambda} \right) \left( \Sigma_{X\hat{X}}^{-1} \Sigma_X - I - \frac{\beta}{2\lambda} \Sigma_Y^{-1} \right)^{-1} \text{,} \label{eqn:vec_gauss_kkt1} \\
\Sigma_{X\hat{X}} &= 2\lambda \Sigma_X (\Sigma_{\hat{X}} - \Sigma_{X\hat{X}}^T \Sigma_X^{-1} \Sigma_{X\hat{X}}).
\label{eqn:vec_gauss_kkt2}
\end{align}
Numerical details and convergence of the algorithm are discussed in Appendix~\ref{appen:vector_gauss_numerics}.

\section{Conclusion}
In this work, we conceptualized the notion of deception with optimal compression highlighting several applications which are consistent with the deception criterion. We defined the rate-distortion-deception (RDD) function and characterized the set of achievable rates that simultaneously satisfy both the distortion and the deception constraints. We then evaluated this RDD function for several commonly occurring random variables such as Bernoulli on the discrete side and Gaussian on the continuous side. Future work involves jointly tackling deception and perception under the same framework together with distortion and exploring task-oriented compression schemes more rigorously.

\appendices
\section{Proof of Lemma \ref{lem:RDP_convex}}\label{appen:lem_RDP_convex}
Let $(D_1,P_1),(D_2,P_2) \in S$. Let $D'=\theta D_1+(1-\theta)D_2$ and $P'=\theta P_1+(1-\theta)P_2$ for $\theta \in [0,1]$. Let $\epsilon>0$. Since $S$ is convex, $(D',P')\in S$ and $R_Y(D',P')$ is well-defined. Since $(D_1,P_1)\in S$, $R_Y(D_1,P_1)$ is well-defined, there exists $p_{\hat{X}|X}$ such that $I_p(\hat{X};X)<R_Y(D_1,P_1)+\epsilon$. Similarly, we have $q_{\hat{X}|X}$ such that $I_q(\hat{X};X)<R_Y(D_2,P_2)+\epsilon$. Now, consider $r_{\hat{X}|X}=\theta p_{\hat{X}|X} +(1-\theta)q_{\hat{X}|X}$. Due to the convexity of the KL divergence and the linearity of the distortion function with respect to the conditional distribution of $\hat{X}$ given $X$, we have that $r_{\hat{X}|X}\in S_{D',P'}$. Therefore, we have,
\begin{align}
    \!R_Y(D',P')\leq &I_{r}(\hat{X};X)\\
    \leq &\theta I_{p}(\hat{X};X)+(1-\theta)I_{q}(\hat{X};X)\\
    \leq &\theta R_Y(D_1,P_1)+(1-\theta)R_Y(D_2,P_2)+\epsilon.
\end{align}
Since $\epsilon$ is arbitrary, we have $R_Y(D',P')\leq \theta R_Y(D_1,P_1)+(1-\theta)R_Y(D_2,P_2)$. Monotonicity of $R_Y(D,P)$ follows from the monotonicity of $S_{D,P}$ with respect to set inclusion.

\section{Proof of Theorem \ref{thrm:achievebility}}\label{appen:thrm_achievability}
First, we will prove the converse of the theorem. We must show that if the triplet $(R,D,P)$ is achievable, then $S_{D,P}\neq\emptyset$ and $R\geq R_Y(D,P)$. Thus, we must first show that there exists a mapping $p_{\hat{X}|X}$ that satisfies the single-letter distortion and deception constraints, if $(R,D,P)$ is achievable through some $n$ letter code. Suppose $(R,D,P)$ is achievable. Then, for some fixed $\epsilon$, there exists an $n$, an $n$-letter encoder $f$, and an $n$-letter decoder $g$, such that \eqref{eqn:n_dist} and \eqref{eqn:n_deception} are satisfied. 

Now, consider the following single-letter encoders $f_i:\mathcal{X}\cross U'\to \mathbb{N}_0$, where for $U'$ we will incorporate the common randomness $U$ and some additional randomness. We encode $x$ using $f_i$ as follows. Draw $n$ i.i.d.~realizations $x'_i$ from $P_X$. Replace $x'_i$ with $x$ and encode the $n$-letter vector using $f$. The unreplaced $x'$ will be our additional source of randomness for the encoder. Now, to decode it, we will use $g$ and extract the $i$th component only. This encoding and decoding scheme defines a stochastic mapping which results in the $D_i=\e[d(X_i,\hat{X_i})]$ distortion and $P_i=D_{KL}(P_{\hat{X}_i}||P_Y)$. Therefore, $S_{D_i,P_i}\neq\emptyset$ and $(D_i,P_i)\in S$, $\forall i$. Since $S$ in convex, we have that $(D'=\frac{1}{n}\sum_{i=1}^nD_i,P'=\frac{1}{n}\sum_{i=1}^nP_i)\in S$. Therefore, $S_{D',P'} \neq \emptyset$. Since $(R,D,P)$ is achievable, we have that $D'\leq D$ and $P'\leq P$. Hence, $S_{D,P}\neq \emptyset$ and $R_Y(D,P)$ is well-defined. Now, to show that $R\geq R_Y(D,P)$,
\begin{align}
    n(R+\epsilon)&>H(f(X^n,U)|U)  \\
    &\geq I(X^n;f(X^n,U)|U)\\
    &= I(X^n;f(X^n,U),U)\\
    &\geq I(X^n;\hat{X}^n)\\
    &\geq \sum_{i=1}^nI(X_i;\hat{X_i})\\
    &\geq \sum_{i=1}^n R_Y(D_i,P_i)\\
    &\geq n\sum_{i=1}^n \frac{1}{n}R_Y(D_i,P_i)\\
    &\geq n R_Y(D',P')\\
    &\geq n R_Y(D,P),
\end{align}
where the equality follows from the fact that $U$ is independent of $X^n$. Therefore, for all $\epsilon>0$, we have $R+\epsilon>R_Y(D,P)$. This proves the desired result. 

Now, to prove the achievability, we will use the strong functional representation lemma \cite{SFRL}. Let $\epsilon>0$. Suppose $S_{D,P}\neq \emptyset$. Therefore, there exits a $p_{\hat{X}|X}\in S_{D,P}$ such that $I(X;\hat{X})< R_Y(D,P)+\frac{\epsilon}{2}$. Now, let $\hat{X}^n$ be generated according to the distribution $\prod_{i=1}^np_{\hat{X}|X}$. By the strong functional lemma, there exists a random variable $Z'$ with support $\mathcal{Z}$ such that $Z'$ is independent of $X^n$, and an encoding function $f':\mathcal{X}^n \cross \mathcal{Z}\to \mathbb{N}_0$ and a decoding function $g':\mathcal{Z}\cross \mathbb{N}_0 \to\mathcal{X}^n$ such that $K=f'(Z',X^n)$ and $\hat{X}^n=g'(Z',K)$. Moreover, we have that,
\begin{align}
    H(f(X^n,Z')|Z')<&I(X^n;\hat{X}^n)+\log( I(X^n;\hat{X}^n)+1)+5 \nonumber\\
    =&nI(X;\hat{X)}+ \log( nI(X;\hat{X})+1)+5.    
\end{align}
This gives us,
\begin{align}
    \frac{H(f(X^n,Z')|Z')}{n}&<I(X;\hat{X})+\frac{\log( nI(X;\hat{X})+1)+5 }{n}\nonumber\\
    &<R(D,P)+\epsilon \leq R+\epsilon,
\end{align}
where $n$ is large enough such that $\frac{\log( nI(X;\hat{X})+1)+5 }{n}<\frac{\epsilon}{2}$. Now, using $Z'$ as our common randomness, $\forall \epsilon>0$, we have an $n$, an encoding function $f'$ and a decoding function $g'$, that satisfy \eqref{eqn:n_dist}, \eqref{eqn:n_deception} and \eqref{eqn:n_entropy}. Therefore, $(R,D,P)$ is achievable.

\section{Simultaneously Diagonalizable $\Sigma_X$ and $\Sigma_Y$ \label{appen:sim_diag_Gauss}}
Assuming zero-mean Gaussians, the Lagrangian is constructed as in (\ref{eqn:lagrangian_vec_gauss}). Since $\Sigma_X$ and $\Sigma_Y$ are simultaneously diagonalizable, we apply the following transformation $\Sigma_{X'}=V\Sigma_XV^T$ and $\Sigma_{Y'}=V\Sigma_YV^T$, where $V$ is an orthogonal matrix ($VV^T=I$) and both $\Sigma_{X'}$ and $\Sigma_{Y'}$ are diagonal. Thus, we are proceeding with this transformation $X'=VX$ and $Y'=VY$. Now, we will compress $X'$ to obtain $\hat{X}'$ which will yield $\hat{X}=V^T\hat{X}'$. Now, the decoupled Lagrangian is,
\begin{align}
    \mathcal{L}(\Sigma_{\hat{X}'},&\Sigma_{X'\hat{X}'},\lambda,\beta) = \frac{1}{2} \ln |\Sigma_{\hat{X}'}| \nonumber\\
    &- \frac{1}{2} \ln |\Sigma_{\hat{X}'} - \Sigma_{X'\hat{X}'}^T \Sigma_{X'}^{-1} \Sigma_{X'\hat{X}'}| \nonumber \\
    &+ \lambda \left( \text{tr}(\Sigma_{X'} + \Sigma_{\hat{X}'} - 2\Sigma_{X'\hat{X}'}) - D \right) \nonumber \\
    &+ \frac{\beta}{2} \Big( \text{tr}(\Sigma_Y^{-1} \Sigma_{\hat{X}'}) -M  + \ln \frac{|\Sigma_Y|}{|\Sigma_{\hat{X}'}|}-P \Big). 
\end{align}
Differentiating the decoupled Lagrangian with respect to $\sigma_{\hat{X}'_i}^2$ and $\sigma_{X_i'\hat{X}'_i}$ gives the KKT equations for each channel $i$,
\begin{align}
    \frac{1}{2\sigma_{X'_i}^2} +\lambda+\frac{\beta}{2\sigma_{Y'_i}^2}-\frac{\beta}{2\sigma_{\hat{X}'_i}^2} &=  \frac{\sigma_{X'_i}^2}{2(\sigma_{X'_i}^2\sigma_{\hat{X}'_i}^2-\sigma_{X_i'\hat{X}'_i}^2)}, \label{eqn:sim_diag_2}\\
    \frac{\sigma_{X_i'\hat{X}_i'}}{\sigma_{X'_i}^2\sigma_{\hat{X}'_i}^2-\sigma_{X'_i\hat{X}'_i}^2} &=2\lambda.\label{eqn:sim_diag_1}
\end{align}
To solve this, we use covariances to write this relation: $\hat{X}'_i = \alpha_i X'_i + Z_i$ where $Z_i \sim \mathcal{N}(0,\sigma_{Z_i}^2)$ and it satisfies $X'_i \perp \!\!\! \perp Z_i$, i.e. independent noise. Thus, $\sigma_{X_i'\hat{X}'_i}=\alpha_i\sigma_{X'_i}^2$ and $\sigma_{\hat{X}'_i}^2=\alpha_i^2\sigma_{X'_i}^2+\sigma_{Z_i}^2$. After substituting these into \eqref{eqn:sim_diag_1} we get the relationship,
\begin{align}
    \sigma_{Z_i}^2=\frac{\alpha_i}{2\lambda}.
\end{align}

\subsubsection{Region 1. Classical Rate-Distortion} In this regime the deception constraint is inactive. Thus, this reduces to the standard Gaussian water-filling solution,
\begin{align}
    \alpha_i&=\max\left(0, 1-\frac{1}{2\lambda\sigma_{X'_i}^2}\right), \\
    \sigma_{Z_i}^2& = \frac{\alpha_i}{2\lambda},\\
    \sigma_{\hat{X}'_i}^2& = \max\left(0, \sigma_{X'_i}^2-\frac{1}{2\lambda}\right).
\end{align}

\subsubsection{Region 2. Zero-Rate Regime} In here, the distortion constraint is inactive. Thus, the system minimizes mutual information $(R_Y(D,P)=0)$ by independently generating $\hat{X}_i'$ as long as it is within the target deception.

\subsubsection{Region 3. The RDD Tradeoff Regime}
Both constraints are active. Substituting $\sigma_{Z_i}^2=\frac{\alpha_i}{2\lambda}$ into  \eqref{eqn:sim_diag_2} gives $A_i\alpha_i^2+B_i\alpha_i+C_i=0$, where
\begin{align}
    A_i = 2\lambda \sigma_{X'_i}^2 + \frac{\beta \sigma_{X'_i}^2}{\sigma_{Y'_i}^2}, &\quad
    B_i = 1 - 2\lambda \sigma_{X'_i}^2 + \frac{\beta}{2\lambda \sigma_{Y'_i}^2}, \nonumber\\
    C_i &= -\beta.
\end{align}
Since we require $\alpha_i\geq0$, we take the positive root,
\begin{align}
    \alpha_i=\frac{-B_i + \sqrt{B_i^2 + 4A_i\beta}}{2A_i}&, \quad
    \sigma_{Z_i}^2 = \frac{\alpha_i}{2\lambda},\nonumber\\
    \sigma_{\hat{X}'_i}^2 = \alpha_i^2 \sigma_{X'_i}^2 + \frac{\alpha_i}{2\lambda}.
\end{align}

\section{Numerical Details and the Convergence for the Vector Gaussian Case Region 3 \label{appen:vector_gauss_numerics}}
For Region 3 of the vector Gaussian case ($\lambda > 0, \beta > 0$), the system does not have a closed-form solution unless $\Sigma_X$ and $\Sigma_Y$ are simultaneously diagonalizable. For the case where $\Sigma_X$ and $\Sigma_Y$ are not simultaneously diagonalizable, we find a  solution by numerically solving the coupled algebraic equations (\ref{eqn:vec_gauss_kkt1}) and (\ref{eqn:vec_gauss_kkt2}) using an alternating minimization approach.

To ensure numerical stability, the algorithm updates the matrices iteratively in a Riccati-like fashion. Given the estimates at iteration $k$, the reconstruction covariance $\Sigma_{\hat{X}}$ is first approximated by computing the pseudo-inverse ($^\dagger$) to avoid near-singularities,
\begin{align}
    \tilde{\Sigma}_{\hat{X}}^{(k+1)} = \frac{1-\beta}{2\lambda} \left( (\Sigma_{X\hat{X}}^{(k)})^{-1} \Sigma_X - I - \frac{\beta}{2\lambda} \Sigma_Y^{-1} \right)^{\dagger}.
\end{align}

This result is then projected onto the positive semi-definite (PSD) cone, $\mathcal{P}_{PSD}(\cdot)$, which enforces valid covariance matrix geometry by clipping any negative eigenvalues to a negligible positive threshold. It should be noted that this projection step could compromise the overall accuracy. To prevent oscillations that are inherent in these coupled equations, a dampening factor $\eta$ is used,
\begin{align}
    \Sigma_{\hat{X}}^{(k+1)} = \mathcal{P}_{PSD} \left( \eta \tilde{\Sigma}_{\hat{X}}^{(k+1)} + (1-\eta)\Sigma_{\hat{X}}^{(k)} \right).
\end{align}

Subsequently, the cross-covariance is updated using the newly computed reconstruction covariance,
\begin{align}
    \tilde{\Sigma}_{X\hat{X}}^{(k+1)} = 2\lambda \Sigma_X \left( \Sigma_{\hat{X}}^{(k+1)} - (\Sigma_{X\hat{X}}^{(k)})^T \Sigma_X^{-1} \Sigma_{X\hat{X}}^{(k)} \right).
\end{align}

Because the cross-covariance $\Sigma_{X\hat{X}}$ must remain symmetric, it is explicitly symmetrized before applying the same dampening factor $\eta$,
\begin{align}
    \Sigma_{X\hat{X}}^{(k+1)} = \eta \left[ \frac{\tilde{\Sigma}_{X\hat{X}}^{(k+1)} + (\tilde{\Sigma}_{X\hat{X}}^{(k+1)})^T}{2} \right] + (1-\eta)\Sigma_{X\hat{X}}^{(k)}.
\end{align}

Convergence is evaluated continuously by tracking the Frobenius norm of the difference between consecutive matrix updates. The algorithm halts gracefully when the matrix differences for both $\Sigma_{\hat{X}}$ and $\Sigma_{X\hat{X}}$ fall below a predefined strict tolerance threshold.

Even though the RDD optimization problem for vector Gaussian sources is convex with respect to $\Sigma_{\hat{X}|X}$, the fixed-point iteration derived from the KKT conditions in \eqref{eqn:vec_gauss_kkt1} and \eqref{eqn:vec_gauss_kkt2} does not guarantee global convergence. The equations used in here resemble Riccati system which can show numerical instability or oscillations due to $\Sigma_{X\hat{X}}^{-1}$ and non-commutativity of $\Sigma_X$ and $\Sigma_Y$. To avoid this, we use a dampening factor $\eta$ and projection onto the positive semi-definite cone $\mathcal{P}_{PSD}(\cdot)$. Empirically, with a suitable $\eta$ this alternating approach is stable and monotonically converges to stationary point that satisfies KKT conditions.

To validate the numerical solution found by the alternating minimization algorithm, we reformulate the Region 3 optimization as a strictly convex log-det program. Since the optimization space over valid covariance matrices is convex solving the primal problem with interior point method (IPMs) gives us a theoretical guarantee of global convergence.

We want to minimize $I(X;\hat{X})$ which is equivalent to maximizing $h(X|\hat{X})$ and as a result maximizing $\ln|\Sigma_{X|\hat{X}}|$. The conditional covariance is given by $\Sigma_{X|\hat{X}} = \Sigma_X - \Sigma_{X\hat{X}} \Sigma_{\hat{X}}^{-1} \Sigma_{X\hat{X}}^T$, to linearize the $\Sigma_{\hat{X}}^{-1}$ we introduce a symmetric positive-definite matrix $W\in\mathbb{S}^M_{++}$ such that $W\preceq\Sigma_{X|\hat{X}}$. By applying the Schur complement, the RDD problem is expressed as this convex program,
\begin{align}
    \max_{W, \Sigma_{\hat{X}}, \Sigma_{X\hat{X}}} &  \ln |W| \quad \nonumber \\
     \st \quad & \begin{bmatrix} \Sigma_{\hat{X}} & \Sigma_{X\hat{X}}^T \\ \Sigma_{X\hat{X}} & \Sigma_X - W \end{bmatrix} \succeq 0 \label{eqn:sdp_schur} \\
    & \text{tr}(\Sigma_X + \Sigma_{\hat{X}} - \Sigma_{X\hat{X}} - \Sigma_{X\hat{X}}^T) \leq D \label{eqn:sdp_dist} \\
    & \text{tr}(\Sigma_Y^{-1} \Sigma_{\hat{X}}) - \ln |\Sigma_{\hat{X}}| \leq 2P + M - \ln |\Sigma_Y| \label{eqn:sdp_perc}
\end{align}
where (\ref{eqn:sdp_schur}) enforces the $\Sigma_{X|\hat{X}}$ lower bound, (\ref{eqn:sdp_dist}) is the distortion constraint, and (\ref{eqn:sdp_perc}) defines the convex feasible set for the deception constraint.

While solving (\ref{eqn:sdp_schur})-(\ref{eqn:sdp_perc}) with IPM solvers guarantees convergence to global optimum, it scales poorly with dimension compared to our alternating minimization approach. Thus, the log-det formulation serves as the theoretical ground truth to compare our alternating minimization algorithm. 

\bibliographystyle{unsrt}
\bibliography{refs}
\end{document}